\documentclass{czjphys}         
\usepackage{amsmath,epsfig,rotating}
\usepackage[english]{babel}
\newcommand{\nn}{\nonumber}
\begin{document}
%
%
\title{Some Analytical Aspects in Field Theories with  Dynamical Mass Generation} 
\authori{Vladim\'{\i}r \v{S}auli}      
\addressi{Dept. Theor. Phys. INP \v{R}e\v{z} near Prague, AVCR}
\authorii{}     \addressii{}
\authoriii{}    \addressiii{}
\authoriv{}     \addressiv{}
\authorv{}      \addressv{}
\authorvi{}     \addressvi{}
\headauthor{V. \v{S}auli}   
\headtitle{Aspects of DMG   \ldots}
\lastevenhead{a to je konec tadydadydaa }
%
\refnum{X}
\daterec{X}    
\issuenumber{1}  \year{1}
\setcounter{page}{1}
\maketitle
%
\begin{abstract}
The consequence of dynamical mass generation on the singularity structure of propagators
is discussed. First the phenomena of dynamical mass generation is discussed in the framework
of Euclidean gap equations, then a possible Minkowski solution is looking for. 
The examples are reviewed and studied for several  models: Yukawa, QED, QCD and Wess-Zumino. 
It is argued that the absence of propagator pole goes hand by hand with the nontrivial solution for mass function. The consequences are discussed. 
\end{abstract}

\section{Introduction}

In perturbation theory the S-matrix elements are composed from Greens functions which are calculated and regularized/renormalized order by order according to the standard rules. By  construction, the correct analyticity of the S-matrix is ensured by the  correct analyticity of Greens functions. As a consequence of assumed analyticity various Greens functions satisfy known integral representations. As a  simple  but  important example  recall the spectral 
representation of fermion propagator
\begin{eqnarray} \label{spek}
S(p)=\int d\omega \frac{\not p\sigma_1(\omega)
+\sigma_2(\omega)}{p^2-\omega+i\epsilon}\quad .
\end{eqnarray}
and the appropriate dispersion relation for selfenergy function:
\begin{eqnarray} \label{sesmek}
\Sigma(p)=\int d\omega \frac{\not p\rho_v(\omega)
+\rho_s(\omega)}{p^2-\omega+i\epsilon}
\end{eqnarray}
where two (in parity conserving theory)  $\sigma's$ are called spectral functions. These are uniquely determined by  knowledge of selfenergy absorptive parts $\rho$'s 
(note $S^{-1}(p)=\not p-m-\Sigma(p)$,  $m$ is a fermion mass and we do not address 
the question of renormalization here). 
In the case of small coupling constant the later functions can be read from  the perturbation theory series of Feynman graphs. In many practical cases the use of perturbation theory is not only very useful, but almost unavoidable tool for meaningful predictions in particle physics.

In this paper we  explore the analytical properties of propagators in strong coupling theories where  nonperturbative approaches are necessary and we deal with  the quantum field models with dynamical mass generation (DMG). For this purpose we consider no mass term of given field in  the  Lagrangian quantum field theory, but assume nontrivial solution  of the mass function. 
In the case of fermion field, the mass function $M$ is conventionally defined as
\begin{equation}
S(p)=\frac{F(p)}{\not p-M(p)} \, ,
\end{equation}
where $F(p)$ is usually called fermion renormalization function. The identification with the previous definitions (\ref{spek}),(\ref{sesmek}) is straightforward.
For a review of strong interacting models with the effect of DMG studied in the context of gap equations see \cite{PAGELS}.

 In this paper we demonstrate on  few model examples
that the correlation function of the field with DMG  does not posses a pole on the real positive semi-axis of complex momenta. In  other words the 'would be particle' associated with the field 
cannot live on-shell. As an hopeful convincing nonperturbative 
tool we use the formalism of Schwinger-Dyson (gap) equations and/or certain 
numerical analysis of some recent lattice QCD data. Recall, the gap equations are exact consequences of quantum theory and they are derivable from the appropriate path integral \cite{SDE}. 

The theories with phenomenon of DMG play an important role in the particle physics. 
First, let us mention the Standard Model. There, the 
dynamical chiral symmetry breaking is clearly inherent (although approximate)
property of light quark sector in QCD, while the  tree or 'bare' masses  are generated via usual Higgs mechanism.  Furthermore, as a very tempting alternatives to the SM  the  Technicolors (or extended Technicolors) models were developed (for a review see \cite{KING}).
These typically Higgsless models do not require non-zero expectation value of scalar field and the electroweak symmetry   $SU(2)_w*U(1)_Y/U(1)_{em}$ is broken dynamically. This is associated with formation of fermion condensates, which is typically driven by a new, possibly non-Abelian strong force and hence it reminds  global chiral symmetry breaking $SU(2)_L*SU(2)_R/SU(2)_V$ in $u,d$ quark's QCD sector. Having no direct experimental evidence for Technicolor scenario the masses of a techniparticles should be scaled up to (at least) several TeV. For a possible indirect experimental evidence and some recent constrains on technifermion models see the papers \cite{APLE2004}, \cite{APPPIASHR2004}. Let us note for completeness here that the role of scalar field and the usual (but now strong) Yukawa interaction can be reconsidered for this purpose  \cite{HAKIOK2005},\cite{HOSEK}.

Further, there is another set of interesting models which suggests the radiative origin of  (at least  the lightest) massive fermions. The so called Zee model \cite{ZEE1980},\cite{CHANZEE1999}  assume the existence of an additional scalar $SU(2)$ doublet with the ordinary Lagrangian mass term. The origin of  small neutrino masses should be economically explained as a pure radiative corrections effect that follows from  the Yukawa coupling constant matrix. Regarding contemporary measurements
of neutrino mixing the Zee model is still one of  attractive models for neutrino mass matrix. 
In order to enhance the reliability of the model at high energy scale, the Zee model has been 
embedded into the supersymmetry \cite{CHEUKON1999}.

\section{Using Standard Methods}

On an example of  the simple model we review or mention the standard methods  usually used in the literature. For this purpose we choose the model already investigated by Bicudo in \cite{BICUDO}. This is in fact simplification of  Schwinger-Dyson equation for fermion in gauge theory and/or in  
the Yukawa model with the (nonhermitian) interaction  $ig\bar{\psi}\psi\phi$  where only the scalar field mass term $m_b^2\phi^2/2$ is explicitly included in the Lagrangian. For further  simplification the ultraviolet divergences are not addressed here and the interaction kernel of gap equation is regularized by a Pauli-Villars term. Having the integral kernel finite, we approximate the field renormalization function, $F=1$ in the equations.
Then the 'mass gap equation' is a non-linear integral equation for the mass function $M$,

\begin{eqnarray}\label{SDEB}
M(p^2)&=&ig^2\frac{Tr}{4}\int\frac{d^4q}{(2\pi)^4}\Gamma(q,l)G_{\Lambda}(p-q)S(q)\, ,
\nn \\
G_{\Lambda}(x)&=&\frac{1}{x^2-m_b^2+i\epsilon}-\frac{1}{x^2-{\Lambda}^2+i\epsilon}
\end{eqnarray}

where $\Gamma$ is a boson-fermion interacting vertex, $G_{\Lambda}$ is  boson propagator  with 
Pauli-Villars regulator. The trace is taken over the dirac indices.
Boson propagator as well as the vertex $\Gamma$ satisfy their own Schwinger-Dyson equations.
For the further technical simplification we take them in their bare form, i.e $\Gamma=1, G^{-1}=p^2-m_b^2$.

Chiral symmetry breaking is associated with a nontrivial solution for a mass function $M$.
Depending on the details of the model it typically appears above a certain critical value 
of the coupling  $ \alpha_c$, where as usual we denote $\alpha=g^2/(4\pi)$. 
Hence to solve the problem of dynamical mass generation is nontrivial task since the kernel of  
eq. (\ref{SDEB}) involves (potentially complex) singularities while
the value of the  coupling, $ \alpha_c\simeq 1$ suffer from the usage of known standard 
perturbative technique. Recall here that   $ \alpha_c=\pi/4$ \cite{BASLOR1999} in 
Euclidean formulation of  presented model when $m_b=0$ and  when a hard cutoff regularization
is used.

Substituting the Ansatz (\ref{spek}) into  eq. (\ref{SDEB})
we get:
\begin{eqnarray}
M(p^2)&=& ig^2\int\frac{d^4q}{(2\pi)^4}
\int d\omega \frac{\sigma_2(\omega)}{p^2-\omega+i\epsilon}G_{\Lambda}(p-q)
\end{eqnarray}
In the standard manner we can obtain the resulting dispersion relation
for the mass function $M$
\begin{eqnarray} \label{pispunta}
M(p^2)&=& \int d\omega \frac{\rho_s(\omega)}{p^2-\omega+i\epsilon}\, '
\nn \\
\rho_s(\omega)&=&\frac{\alpha}{(4\pi)}\int d x \sigma_2(x) \left[X_0(\omega;m_b^2,x)-
X_0(\omega;\Lambda^2,x)\right]\, ,
\nn \\
X_0(\omega;a,b)&=&\frac{\sqrt{(\omega-a-b)^2-4ab)}}{\omega}
\Theta\left(\omega-(\sqrt{a}-\sqrt{b})^2\right)\, ,
\end{eqnarray}
where $\lambda(\omega,a,b)$ is  well-known  triangle function
$$\lambda(\omega,a,b)=(\omega-a-b)^2-4yz=\omega^2+a^2+b^2-2\omega a-2\omega b-2ab$$.

The imaginary part of propagator and the absorptive part of selfenergy $M$ are related through
the following {\it Unitary equation}\cite{LACO}, 
\begin{eqnarray} \label{UE}
\sigma_2^c(p^2)&=&\frac{\rho_s(p^2)}{c_s(p^2)}+\frac{c_v^2(p^2)}{c_s^2(p^2)}p^2\sigma_2(p^2)
\\
c_s(p^2)&=&p^2+Re M^2(p^2)+\pi^2\rho_s^2(p^2)
\nn \\
c_v(p^2)&=&-2Re M(p^2)\, .
 \nn
\end{eqnarray}
which, as can be deduced from the Heaviside step functions in eq. (\ref{pispunta}), are 
nonzero above the threshold.
In eq. (\ref{UE}) the above index $c$ means the continuous part of the spectral function. 
From its derivation it follows that 
the eq. (\ref{UE}) requires  the principal value integration.
This can be performed in analytical closed form:
\begin{eqnarray}
Re M(p^2)=P.\int d\omega \frac{\rho_s(\omega)}{p^2-\omega}=
\int d x \sigma_2(x)\left[J(p^2,x,m_b^2)-J(p^2,x,\Lambda^2)\right]\, ,
\end{eqnarray}
where we have labeled
\begin{eqnarray}
J(p^2,y,z)&=&
-\frac{\Theta(-\lambda_p)\sqrt{-\lambda_p}}{p^2}
\left[\frac{\pi}{2}+arctg\frac{p^2-y-z}{\sqrt{-\lambda_p}}\right]+\frac{1}{2}\ln(16yz)
\nn \\
&-& \frac{\Theta(\lambda_p)\sqrt(\lambda_p)}{p^2}
\ln\left|\frac{p^2-y-z+\frac{\lambda_p}{T-p^2}}{p^2-y-z+\sqrt{\lambda_p}}\right|
+\frac{\Theta(\lambda_0)\sqrt{\lambda_0}}{p^2}
\ln\left|\frac{-y-z+\frac{\lambda_0}{T}}{-y-z+\sqrt{\lambda_0}}\right|\, ,
\nn \\
\end{eqnarray}
where we have introduced the following abbreviations:
\begin{eqnarray}
\lambda_p&=&\lambda(p^2,y,z) ,
\nn \\
\sqrt{\lambda_0}&=&|y-z|\, ,
\nn \\
T&=&(\sqrt{y}+\sqrt{z})^2\, .
\end{eqnarray}
Note that having $m_b$ different from zero one can exclude the possibility  that the spectral 
function $\sigma$ is   purely continuous. 
The inverse of propagator becomes complex from the threshold
$(T+m_b)^2$ while $T$ is a lower boundary of the spectral integral (\ref{spek}), i.e the point where the propagator becomes complex. This is clear contradiction and the  Unitary Equations  have no solution in this case. Hence we assume  usual following form 
\begin{eqnarray}
\sigma_2(x)=r\delta(x-m^2)+\sigma_2^c(x)
\end{eqnarray}
where $r$ is a propagator residuum and $m$ is a pole mass. These two propagator characteristics
can be selfconsistently determined as follows
\begin{eqnarray}
r&=&\frac{1}{1+2m\int\frac{\rho_s(x) dx}{(m^2-x)^2}}\, 
\nn \\
m&=&\int\frac{\rho_s(x) dx}{m^2-x}\, .
\end{eqnarray}

As the second and in fact most spread technical method of solution we make our calculation in the Euclidean space. In the Euclidean formalism the  analyticity domain of propagator functions should be the same as in the previous spectral-Minkowski treatment. Assuming no poles in the first and third quadrants of complex momenta one 
can deform the integration contour and perform the so-called Wick rotation $iq_0\rightarrow q_4$, where $q_4$
is the fourth component of Euclidean momenta $q^2_E=q^2$(and similarly for $p$).
 Choosing then convenient angular parametrization
\begin{eqnarray}
q_1&=&q_E\sin\phi\sin\theta\sin\eta 
\nn \\ 
q_2&=&q_E\cos\phi\sin\theta\sin\eta
\nn \\
q_3&=&q_E\cos\theta\sin\eta
\nn \\
q_4&=&q_E\cos\eta
\end{eqnarray}
such that $(p_E-q_E)^2=p_E^2-2p_Eq_E\cos\eta +q^2_E$
one can integrate over  all angles which leave us  with only one dimensional integral equation for the function $M$
\begin{eqnarray} \label{euclid}
M(p^2)=\frac{\alpha}{2.4\pi p^2}\int_0^{\infty}d q^2_E \, M(q^2_E)
\frac{-m_b^2+\Lambda^2+\sqrt{\lambda(p^2_E,q^2_E,m_b^2)}-\sqrt{\lambda(p^2_E,q^2_E,\Lambda^2)}
}{q^2_E+M^2_E(q^2)}
\end{eqnarray}
(see also \cite{BICUDO}). The equation (\ref{euclid}) 
can be easily solved numerically by the method of iteration.

Assuming uniqueness of the spectral representation for fermion propagator
and using the theorem of uniqueness of the analytical 
continuation we should find that the function $M(q^2)$ obtained as a solution of the Unitary Equations (\ref{UE}) 
 must be also the solution of the equation (\ref{euclid}) for $M(q^2_E)$ at spacelike momenta.  
The  disagreement implies the 'unexpected' position of Greens functions 
singularity. We should stress here that the spectral  and the Euclidean solutions  actually coincide in weak coupling quantum field models  with explicit mass presented  \cite{LACO}.

Before the presentation of the results let us mentioned further 
possible method of  solving gap equations in Minkowski space.
This is based on  a partial neglecting of  the time momentum component
in the kernel of the equation. Typically  only the assumed dominant part of 
propagator poles is  considered. These and similar approaches were mostly used 
in the calculation of bound state equation, while  for a given model some review can be found in  \cite{BICUDO}. However, the dynamical chiral symmetry breaking is deeply nonperturbative -strong coupling- phenomenon and 
we believe it is difficult to have such an approximations under  safe control and  we do not employ this method. 
\begin{figure}
\centerline{\epsfig{figure=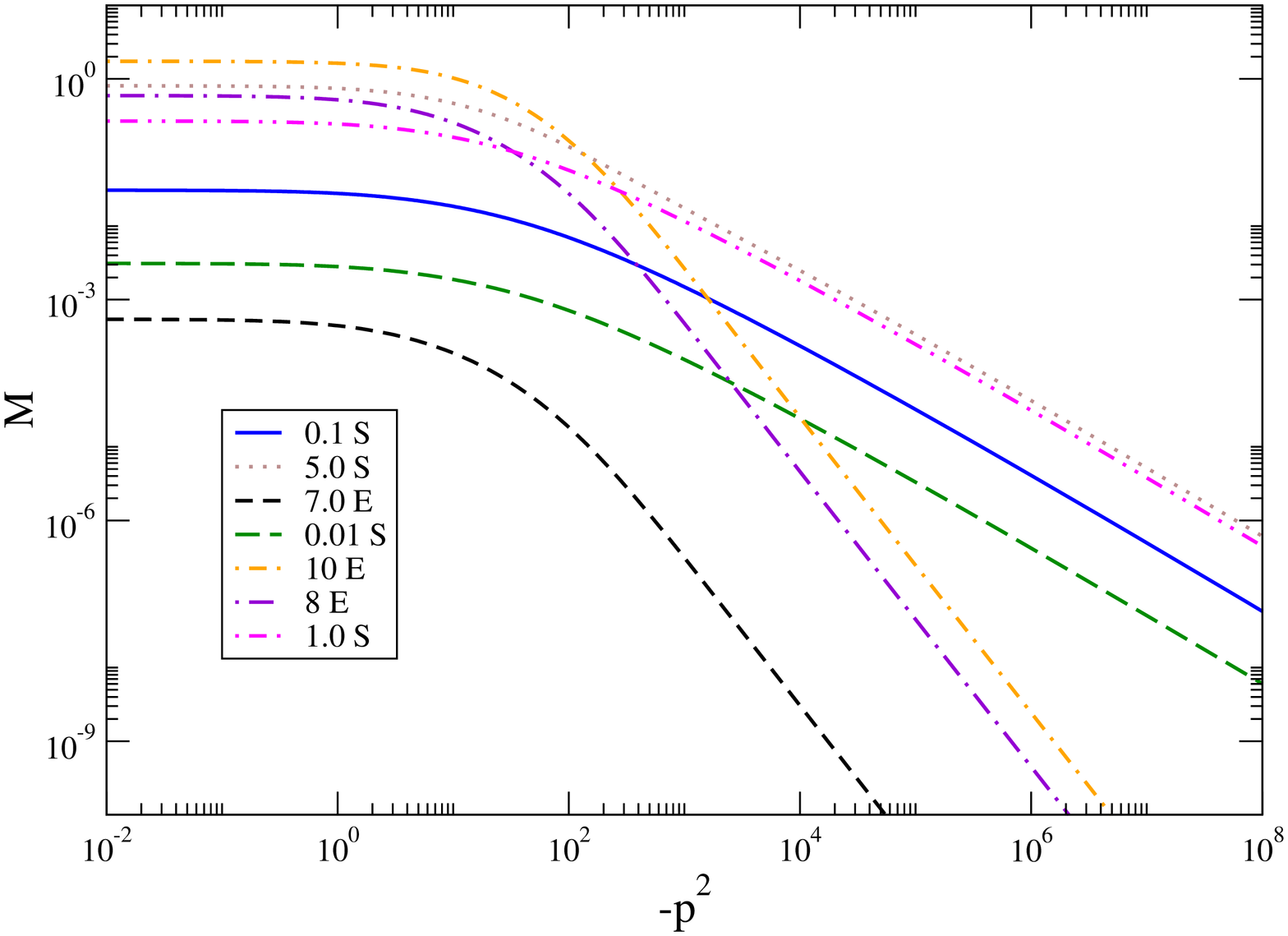,width=8truecm,height=8truecm,angle=270},
\hspace*{-1.0truecm},
\epsfig{figure=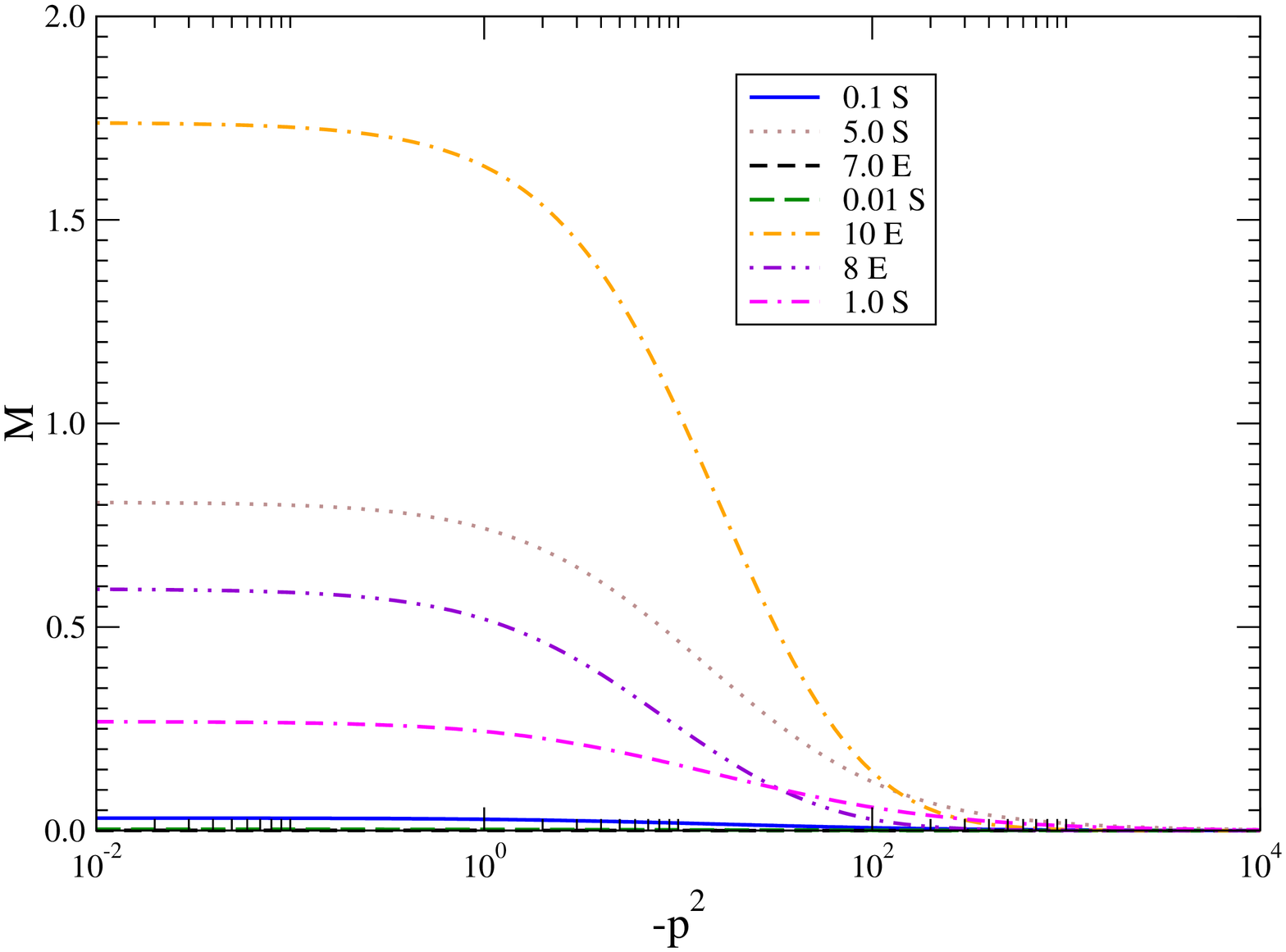,width=8truecm,height=7truecm,angle=270}}
\caption[caption]{Dynamical mass as obtained in the Euclidean (E) formalism and spectral (S) formalism. Left (right) panel shows the result in log-log (log-linear) axis scale. Each line is labeled by the coupling constant. }
\label{jukavy}
\end{figure}
In what follows the numerical iteration of the equations offers stable convergent and accurate solutions in both approaches. We take the boson mass $m_b=1$  and all the results are scaled in these units   in our actual calculations. The square of the cutoff is $\Lambda^2=50$. In the Euclidean theory such constellation  leads to the expected dynamical chiral symmetry for the coupling  larger then $\alpha_c\simeq 6$. The infrared value  of generated mass, say $M(0)$, has a drastic  dependence on the coupling constant. We did not attempted to parametrize this behaviour but recall here the paper \cite{BASLOR1999} where, within a zero boson mass  the authors of  \cite{BASLOR1999}
have found  the well-known exponential behaviour (this so called  Miransky scaling is explicitly 
shown in the next section). The appropriate results for the function $M(p^2_E)$ are displayed in fig.1. In what concerns the behaviour of the mass at ultraviolet, Bicudo \cite{BICUDO} shows that
\begin{eqnarray}
M(p^2)\rightarrow\frac{(\Lambda^2-m_b^2)<\bar{\psi}{\psi}>}{p^2}^2\, ,
\end{eqnarray}
where the  fermion condensate is defined as usual $<\bar{\psi}{\psi}>$=gTr S.
This power behaviour seems to be confirmed by our numerical analyzes.

Unfortunately, this is not what we do  observe from the solution of the unitary equations.
First of all let us mentioned that there is no nontrivial critical coupling. 
Stable solution for the functions $\sigma_c$ $(\rho_s)$, residuum $r$ and pole mass $m$ 
is obtained for any non-zero coupling constant $\alpha$.
The mass function for 
spacelike $p^2$ is then calculated through the appropriate dispersion relation and 
compared with $M(p^2_E)$. These results are added to the fig. 1 for several values of coupling constant. From this we can see that these two solutions do not agree for the same coupling. Furthermore, the scaling $M(0)$ is approximately linear for our 'would be' Minkowski solution. Also the  ultraviolet  behaviour 
differs, the 'spectral solution' for $M$  is driven by the first power of the inverse square of momentum, i.e. $M(p^2)\rightarrow const/p^2$. Note, the  possibility that this is  also a solution of   (  non-unique for a while) the gap equation (\ref{euclid}) has been excluded by  simple substitution and comparison. 

Using analyticity arguments above we can conclude that the structure of correlation functions 
at the complex  momentum plane prohibits naive backward Wick rotation $ip_4\rightarrow p_0$ and the propagators does not posses naively expected spectral (Lehmann) representation. Here we stop
our discussion and leave the correct Minkowski treatment for a future interesting investigation.


\section{Another example: Ladder QED}

Considering {\it explicitly massive} electron, the analytical structure  of strong coupling
QED electron propagator was subject of the initial study \cite{FUKKUG1972}. It was found 
that above the certain value of the coupling constant the branch point of the fermion propagator disappears from the real timelike semi-axis. This was also confirmed by the later author's study
made in the paper \cite{SAULIJHEP}. Furthermore, based on the solution of {\it Unitary Equations} presented in the previous section, it was argued that spectral representation (\ref{spek})
is absent in these circumstances (for a further  review of various Euclidean solutions in QED and for a review of similar spectral techniques see the list of  references in \cite{LACO},\cite{PHD}). It is very interesting fact that the above mentioned value of the coupling is exactly identical with the critical coupling $\alpha_c=\frac{\pi}{3}$ which is true at least in the ladder approximation of Schwinger-Dyson equation for $S$.

Remind at this place the trivial fact that the existence of free  particle moving is necessary associated with a presence of    propagator pole at some momentum square $p^2=m^2_p>0$. Hence, if  no fermion mass term is considered in the  Lagrangian and simultaneously  free massive fermions are expected then the naive perturbative dispersion law $E=\sqrt{\vec{p}^2}$ must be nontrivially transformed into $E=\sqrt{\vec{p}^2+m_p^2}$. Here the 'gap' is naturally  identified with the pole mass $m_p$, which  must be a solution of our gap  equation.
On the other hand the absence of a real propagator pole could be a hint for a confinement or 
resonant behaviour.

From the above arguments one can  conjecture that the  confinement 
is somehow intimately related with the dynamical chiral symmetry breaking.
At the introductory section we enumerate several models that propose the dynamical origin of the
particles in the nature. Clearly, having the leptons unconfined it is desirable to look for an interaction which spawn DMG but avoid a confinement. In a gauge theory, the behaviour of the running coupling associated with the proposed new strong interaction plays the key role.
As an simple example let us consider  massless gauge theory where the coupling is not running but is approximated by a constant value.
For further simplification the fermion-antifermion-photon vertex is  approximated by the bare vertex $\gamma^{\mu}$.  In this so called Ladder  
approximation the fermion gap equation reads:
\begin{equation}
\Sigma(p)=\not p-ie^2\int\frac{d^4k}{2\pi^4}G^{\mu\nu}_{0}(k-p)\gamma_{\mu}S(p)\gamma_{\nu}
\end{equation}
where $G^{\mu\nu}_{0}$ is a free massless gauge boson propagator in Landau gauge.

Recall again the most conventional strategy: after the   Wick rotation and the
 angular integration the gap equation in  Euclidean space is formulated:
\begin{equation}
M(p_E)=m_0+\frac{3\alpha}{4\pi}\int_0^{\Lambda^2} dk_E^2 \frac{2k_E^2}{p_E^2+k_E^2+\sqrt{k_E^2-p_E^2}}
\frac{M(x)}{k_E^2+M^2(k_E)}
\end{equation}
where $m_0=0$ when studying chiral symmetry breaking and where we have introduced an Euclidean cutoff regulator $\Lambda$ as the upper boundary of Euclidean integral. Note that now, contrary to previous case, $F=1$ as the exact
solution because of  ladder approximation.

Like in the previous section the second  
alternative possibility is to assume spectral representation (\ref{spek}) and to solve the
Unitary Equation directly in Minkowski space.
The evaluation of selfenergy and its absorptive part $\rho_s$ is straightforward (for details see for instance \cite{LACO}. Note only that the eq. (\ref{UE}) is exact now since $\rho_v=0$ in ladder approximation.

As a first step, mainly due to the pedagogical reasons, we exhibit the agreement of   both mentioned approaches and present solutions for small (in fact subcritical) couplings in an explicit chiral symmetry breaking, i.e. for non-zero $m_0$. For this purpose we prefer to show the renormalized results  obtained by the manner firstly used in the paper \cite{SAULIJHEP}. 
 The 'spectral' solutions at the time-like regime are displayed in Fig.1a. 
The maximum cusps in the real parts of dynamical mass function $M(p)$ correspond to the thresholds, i.e.  to the physical mass of the electron. The continuous imaginary part of $M(p)$, i.e. $\pi\rho_s(p^2)$ is  the result of the solution of  Unitary equations. These are  displayed  in Fig. 1 (left panel). The comparison of Euclidean and spectral solutions  is shown at the right panel of Fig. 1.  Solid line curves represent the Euclidean solutions, the dot-dashed lines are the mass functions obtained via spectral treatment. The coupling constants $\alpha=e^2/4\pi$ decrease from top to bottom. For completeness note that  zero momentum subtraction renormalization scheme was used. At this place we should stress that only the pole mass, defined as  $m_p=\Sigma(m_p)$, is the observable ( it should be  independent on the gauge and on the renormalization scheme).    
\begin{figure}
\hspace*{1.5truecm}\centerline{\epsfig{figure=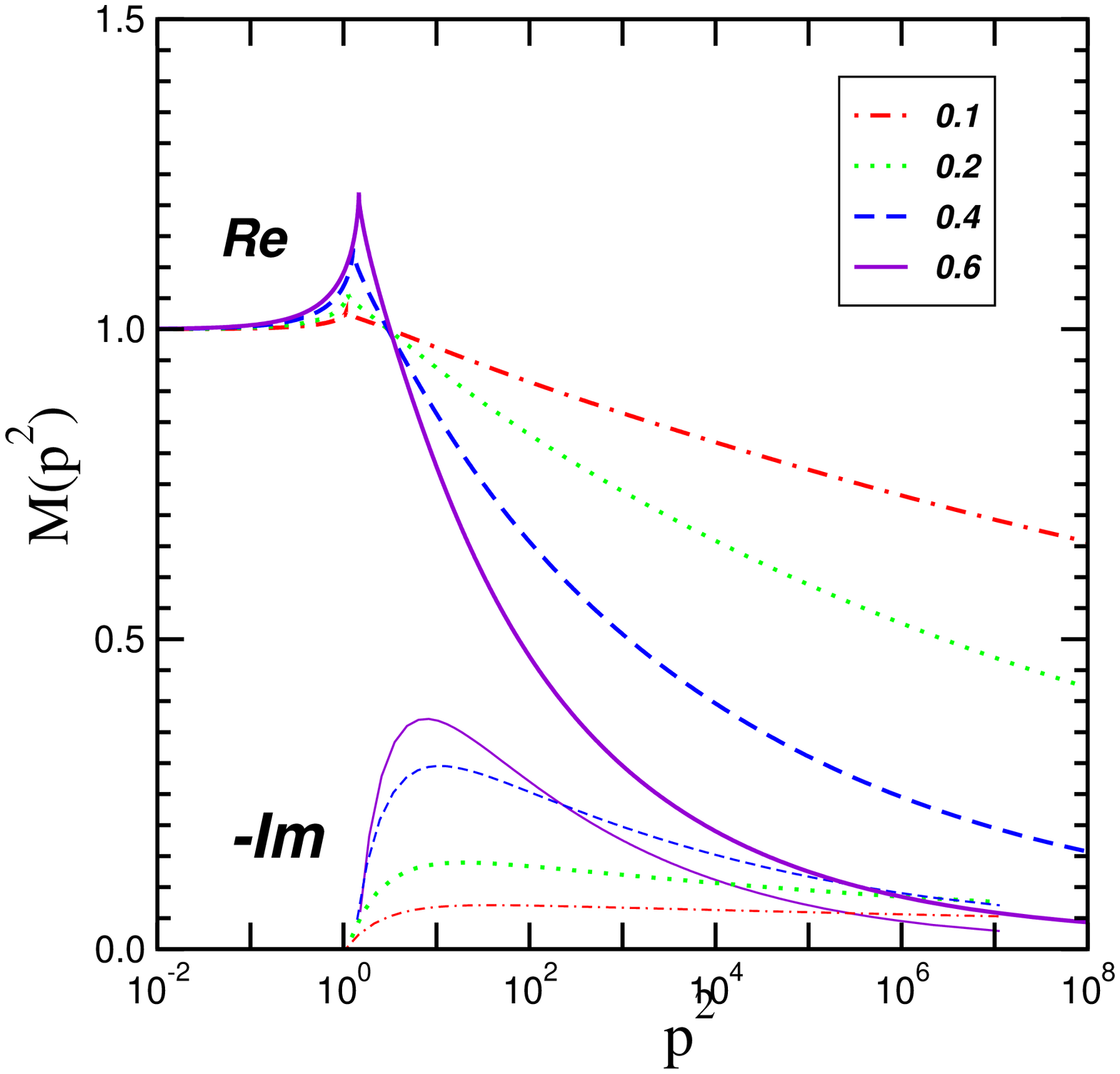,width=8truecm,height=10truecm,angle=270},
\hspace*{-3.7truecm},
\epsfig{figure=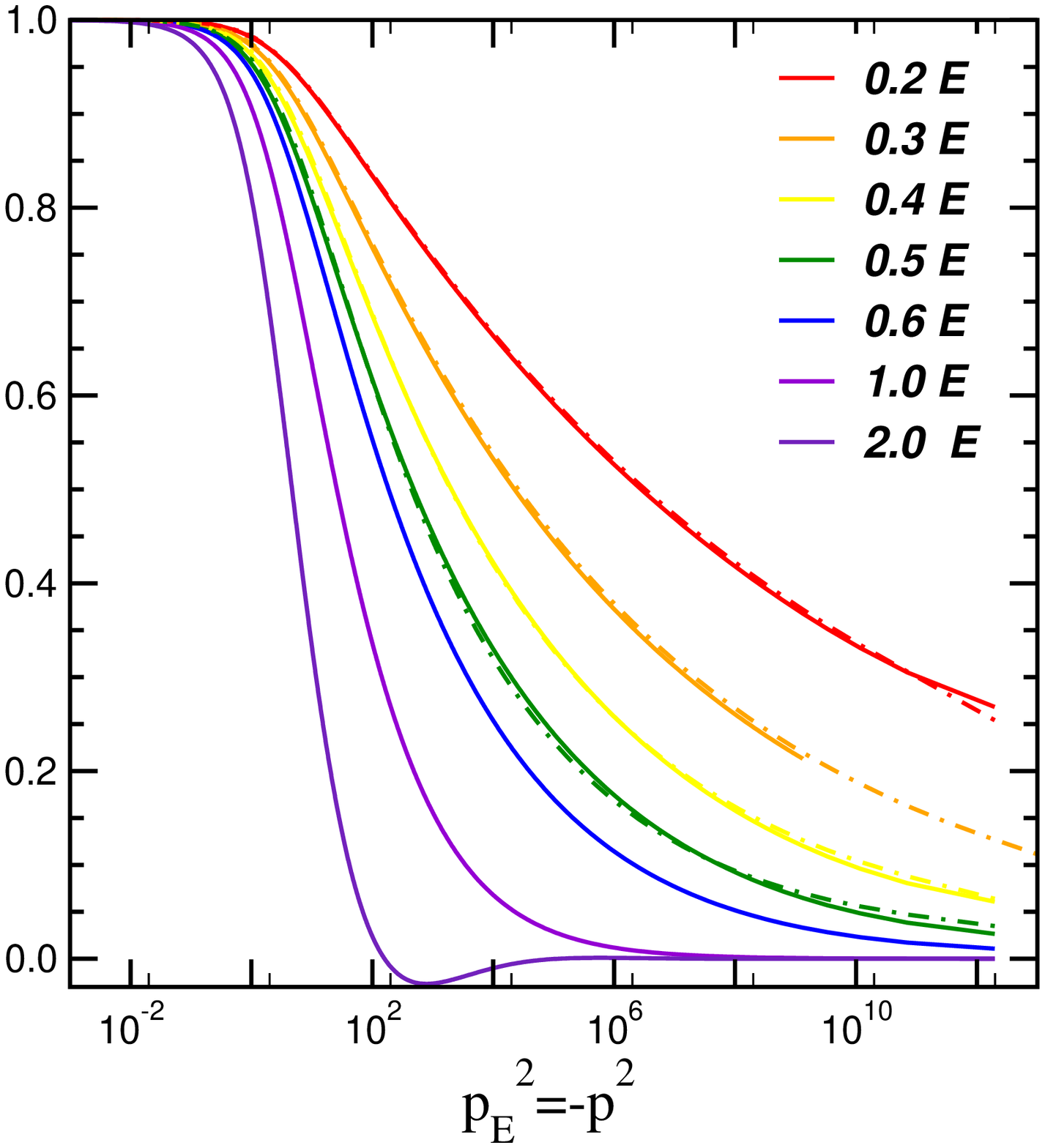,width=8truecm,height=10truecm,angle=270}}
\caption[caption]{Fermion  mass functions in the case of explicit chial symmetry breaking at timelike (left panel) and spacelike (right panel) momenta as described in the text.}
\label{explicit}
\end{figure}
In addition we switch off the electron mass in the  Lagrangian.
In the Euclidean formalism the properties of DMG are well known in this model. 
The mass function in the infrared obeys the Miransky scaling \cite{MIRA}
\begin{eqnarray} \label{miran}
M(0)\sim\Lambda e^{C\sqrt{1-\frac{\alpha_c}{\alpha}}}
\end{eqnarray}
for $\alpha>\alpha_c=\frac{\pi}{3}$, while for 
 $\alpha<\alpha_c$ chiral symmetry is preserved 
and $S=\frac{1}{\not p}$ exactly.

Dynamically generated mass  is displayed in the figure 2 (solid lines). The appropriate curves are  labeled by the coupling constant $\alpha$. The results for dynamical mass $M$ in unquenched QED are added  for interest to. The polarization effect were taken selfconsistently into account and we have used the bare vertex in this case. The Schwinger-Dyson equations in unquenched QED have been solved in several papers \cite{BLOCH,LACO,RAKOW,PHD}. The results presented here follows the numerical treatment described in
\cite{LACO}.
\begin{figure}
{\epsfig{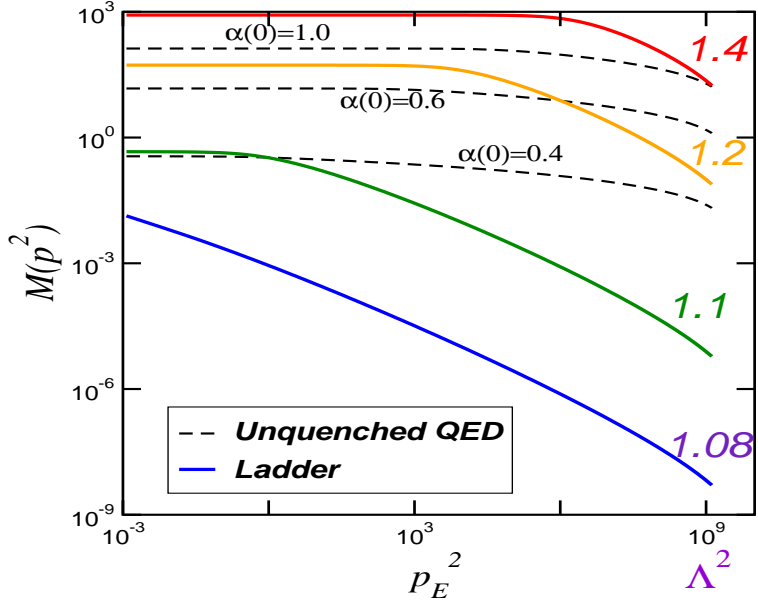}}
\caption[caption]{Dynamical masses in  ladder QED (solid lines) and beyond.}
\label{genmass}
\end{figure}
  What is the timelike behavior? 
It was already shown  by Fukuda\&Kugo  in 1976 \cite{FUKKUG1972}
that there is no evidence for the propagator pole in ladder QED. The function $p^2-M(p^2)$ never cuts the zero axis for  $\alpha>\alpha_c$. This observation is valid even for  explicit chiral symmetry breaking case and it  has been confirmed in the paper 
\cite{SAULIJHEP} again. Yet more, no evidence for solution of our unitary equations  at supercritical regime $\alpha>\alpha_c$ was found (the same is observed in unquenched case). 
It is probably strong hint (but of course not a proof) for a changes of propagator's analyticity domain and we argue that fermion propagator does not posses usual spectral representation in Ladder QED. We do not investigate the analytical structure of the fermion propagator in ladder QED further. Recall, that  one can expect some but not radical influence from vertex corrections which was neglected in our study.

\section{QCD}

The Quantum Chromodynamics  is the only experimentally studied
strongly interacting relativistic quantum field theory. This
non-Abelian gauge theory with a gauge group $SU(3)$ has many
interesting properties. The dynamical spontaneous breaking of chiral
symmetry explains why the pions are light, identifying them with the
pseudo-Goldstone bosons associated with the symmetry breaking of the
group $SU(2)_L\times SU(2)_R$ to $SU(2)_V$ (in flavor space).
Asymptotic freedom \cite{GROWIL1973,POL1973} implies that the
coupling constant of the strong interaction decreases in the
ultraviolet region. For less than 33/2 quark flavors the QCD at high
energy becomes predictable by the perturbation theory. 
However, in the infrared region
the Perturbation theory does not work and non-perturbative 
techniques have to be
applied.

One of the most straightforward non-perturbative approaches is a
solution of the gap equations for QCD. The extensive studies were
undertaken for a quark gap equation, based on various model assumptions for
a gluon propagator. These approximate solutions, often accompanied
by a solution of the fermion-antifermion Bethe-Salpeter for meson states,
have become an efficient tool for studies of many non-perturbative
problems, e.g., the  chiral symmetry breaking, low energy
electroweak hadron form factors, strong form factors of exclusive
processes, etc (see reviews
\cite{ROBERTS,ALKSMEK2001,MARROB2003,ROBSMI2000} and also recent
papers \cite{FISALK2003,BENDER,BICEST2003}).

However, to take gluons into account consistently  is much more
difficult than to solve the quark SDE alone. 
At this place we briefly review the
symmetry preserving {\it gauge invariant} solution obtained by
Cornwall two decades ago \cite{CORN1982}. To our knowledge this is the
best published example, in which the behavior of the QCD Green's
functions in the whole range of Minkowski formalism is addressed
within the framework of the gap equations.

Cornwall reveal the resulting gluon propagator which is  
gauge invariant by the construction and  the gluon form factor 
does not depend on the gauge fixing parameter, although it was used at the beginning of 
the author's calculation. In his approach the  Green's functions satisfy the naive Ward identities which is a direct consequence of {\it gauge }  and {\it pinch technique} introduction.
For the details see the original paper \cite{CORN1982}. It was shown that the resulting gluon propagator can be rather accurately fitted in spacelike regime as     
\begin{eqnarray}
G^{\mu\nu}(q)_{AB}&=&\delta_{AB}G(q)\left(-g^{\mu\nu}+\frac{q^{\mu}q^{\nu}}{q^2}\right)
\nn \\
G^{-1}(q)&=&[-q^2+m^2(q^2)]bg^2ln\left[\frac{-q^2+4m^2(q^2)}{\Lambda^2}\right]
\label{cornwall} \\
m^2(q^2)&=&m^2\left[\frac{ln\left(\frac{-q^2+4m^2}{\Lambda^2}\right)}
{ln\frac{4m^2}{\Lambda^2}}\right]^{-12/11}\, ,
\nn
\end{eqnarray}
where $g$ is a gauge coupling,
$b$ is one loop renormgroup coefficient, $m\simeq [1.5-2] \Lambda$,  
$\Lambda$ has the meaning of $\Lambda_{QCD}$.
The Latin (Greek) letters stand for a color (Lorentz) indices.
One should mention here that this propagator was successfully used in some
recent phenomenological calculations of $\tau$ decay, meson form factors, etc..
, see for instance \cite{AGMINAph,AGMINA2002,MINA2000}.

In recent papers
\cite{FISALK2003,BLOCH2003,ATKBLO1998,KONDO2003,AGUNAT2004} studies
of the coupled SDEs for gluon and ghost propagators in  the Landau
gauge in Euclidean space were performed in various approximations.
However the non-Abelian character of the QCD makes it difficult to convert
the momentum SDEs into equations for spectral functions. We were not
able to do this yet since
the Fadeev-Popov ghosts have to be included \cite{FADPOP1967} in a class
of Lorentz gauges. The main obstacle is the ghost SR, mainly due to
singular behavior at zero momenta 
(see for instance \cite{FISALK2003}).

Hence, instead of solving the SDEs 
the following
generalized spectral representation 
\begin{equation} 
G(q^2)=\int\limits_{0}^{\infty} d\omega\,
\frac{\sigma(\omega;g(\xi),\xi)}{q^2-\omega+i\epsilon} \, ,
\label{gluesp} 
\end{equation}
was used  \cite{LACO} to fit the spectral function
to Euclidean solutions obtained in recent lattice simulations and in
the gap equations formalism for gluon form factor $G$. 
The  full dependencies of the {\it continuous} function
$\sigma(\omega;g(\xi),\xi)$ are indicated explicitly in the  Ansatz (\ref{gluesp}). 
 $g(\xi)$ is the renormalized QCD coupling constant at the scale $\xi$ and the word 'generalized'
simply means that we do not assume (and do not obtain) the
spectral function $\sigma(\omega)$ positive for all values of
$\omega$. 
\begin{figure}
\centerline{
\epsfig{figure=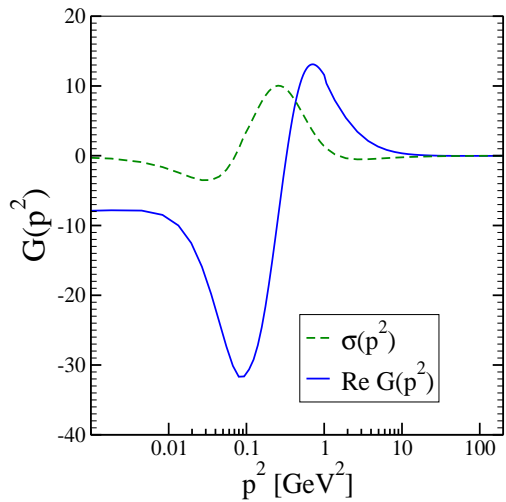,width=8truecm,height=10truecm,angle=270}}
\caption[caption]{ \label{gluontime}Gluon propagator for
timelike momenta.  }
\end{figure}
The  spectral
decomposition has been  fitted to some recent lattice
\cite{BOBOLEWIZA2001} and   SDEs results \cite{FISALK2003} (both
defined in spacelike region) and
then predicts the gluon propagator in the timelike region. 
For a details of the numerical procedure see the original paper
\cite{LACO}.
The solutions are plotted in Fig.~\ref{gluontime}. The spectral function
has a smooth peak around  $p^2=(0.7{\rm GeV})^2$ with the width
$\approx 1$GeV.  It becomes negative for asymptotically large $p^2$,
as  expected already  from the PT. The gluon propagator should
describe the confined particle, so the ``unusual'' shape (violation
of  the spectral function positivity) is in accord with our physical
expectations.

\section{Wess-Zumino model}

At the end of this paper we investigate the phenomenon of dynamical mass generation in the supersymmetrized version of Yukawa interaction. This well-known Wess-Zumino model (WZM) \cite{WESZUM1974} has been already studied in the formalism of gap equations in couple of papers \cite{BASLOR1999,SAUsup}.
In both cases the authors preferred to work with the right number of degrees of freedom in the action and we follow this approach for a convenience. 
The  auxiliary fields $F,G$ of the Wess-Zumino supermultiplet $\phi=(A,B,F,G,\psi)$ were 
integrated out and thus we are left with the following generating functional of WZM:
\begin{eqnarray} \label{geneWSM}
Z_{WZ}[J]&=&\frac{1}{Z[0]}\int {\cal{D}}A{\cal{D}}B{\cal{D}}\psi
\exp\left\{\frac{i}{\hbar}\int d^4x \left[ {\cal{L_{WZ}}}+J_A.A+J_B.B-\bar{\eta}\psi\right]\right\}
\nn \\
L_{WZ}&=&L_{free}+L_{int}   
 \\
L_{free}&=&\frac{1}{2}\left[\partial_{\mu}A\partial^{\mu}A
+\partial_{\mu}B\partial^{\mu}B
+i\bar{\psi}\not\partial\psi
-m^2A^2-m^2B^2-m\bar{\psi}\psi\right]
\nn \\
L_{int}&=&-mgA(A^2+B^2)-\frac{g^2}{2}(A^2+B^2)^2
-g\bar{\psi}\psi A-ig\bar{\psi}\gamma_5\psi B
\nn
\end{eqnarray}
where majorana fermion $\psi$ interacts through the usual Yukawa interaction with scalar
$A$ and pseudoscalar $B$. Since the whole interaction is dictated by the exact
global supersymmetry there should not be quadratic divergences. It allows us to consider WZM as a natural (stable against higher corrections) low energy effective theory.

Here we did not attempt to solve gap equation directly in Minkowski space, but apply the method
first mentioned in the previous section. We will  solve the gap equation in Euclidean space and then, by the means of spectral representation,  we make an analytical continuation to the Minkowski space. Like in the previous study of gluon propagator it is based on a numerical fit of assumed spectral representation at Euclidean (spacelike) domain of momenta.  
  
In the approximation used in the paper \cite{SAUsup} the  gap equations in Euclidean space read:
\begin{eqnarray}
&&\Pi_A(y)=\frac{\alpha}{2\pi^2}\int_0^{\infty} dx\, x
\int_0^{\pi}d\theta \sin^2 \theta{\cal{J_A}}
\nn \\
&&\Pi_B(y)=\frac{\alpha}{2\pi^2}\int_0^{\infty}  dx\, x
\int_0^{\pi}d\theta \sin^2\theta{\cal{J_B}}
\nn \\
&&{\cal{J_A}}=\biggl\{6 G_A(x)+2 G_B(x)
-8\frac{{\cal{F}}(x){\cal{F}}(z)}{x+{\cal{M}}^2(x)}
+4\frac{{\cal{F}}(x){\cal{F}}(z)[y+({\cal{M}}(x)+{\cal{M}}(z))^2]}
{(x+{\cal{M}}^2(x))(z+{\cal{M}}^2(z))}\biggr\}
\nn \\
&&{\cal{J_B}}=\biggl\{6 G_B(x)+2 G_A(x)
-8\frac{{\cal{F}}(x){\cal{F}}(z)}{x+{\cal{M}}^2(x)}
+4\frac{{\cal{F}}(x){\cal{F}}(z)[y+({\cal{M}}(x)-{\cal{M}}(z))^2]}
{(x+{\cal{M}}^2(x))(z+{\cal{M}}^2(z))}\biggr\}
\nn
\end{eqnarray}
\begin{eqnarray}
&&\frac{1}{{\cal{F}}(y)}=1+\frac{2\alpha}{\pi^2}\int_0^{\infty} dx 
\int_0^{\pi}d\theta x\sin^2\theta
\biggl\{ \frac{{\cal{F}}(l)\sqrt{(x/y)}\cos\theta}{x+{\cal{M}}^2(x)}
\left[G_A(z)+G_B(z)\right]\biggr\}
\nn \\
\label{dysoni}
&&\frac{{\cal{M}}(y)}{{\cal{F}}(y)}=\frac{2\alpha}{\pi^2}\int_0^{\infty} dx 
\int_0^{\pi}d\theta x\sin^2\theta
\biggl\{ \frac{{\cal{F}}(x){\cal{M}}(x)}{x+{\cal{M}}^2(x)}
\left[G_A(z)-G_B(z)\right]\biggr\}
\end{eqnarray}
where $x,y$ stands for the square of Euclidean four-momenta, $z=y-2\sqrt{xy}\cos\theta+x$, and the scalar propagators are defined like
$G_{A,B}(x)=[x+\Pi_{A,B}(x)]^{-1}$ and  being interested in the effect of dynamical mass generation we put $m=0$. Note that  all scalar fields cubic  interactions vanish in this case. For interested reader the massive case is reviewed and solved in \cite{SAUsup}.

Here, we instead of the use of hard momentum or Pauli-Villars cutoff  
we put higher derivative cutoff function into work. For simplicity
the following  form of cutoff function: 
\begin{eqnarray} \label{cut}
dx\rightarrow dx \frac{1}{(1+x/\Lambda)(1+z/\Lambda)}
\end{eqnarray}
has been introduced into the  equations written above. The simple  cutoff function
(\ref{cut}) has been chosen  for the purpose of convenience and we assume that any other reasonable choice (including Pauli-Villars regulator) gives the same 
results for $p^2<<\Lambda^2$.

There is only one scalar function characterizing scalar propagator. 
Thanks to (observed) positivity of the selfenergy $\Pi_{A,B}$, 
the Euclidean mass function can be defined like 
$M_{A,B}(p)=\sqrt{\Pi_{A,B}}$. 
The equations (\ref{dysoni}) have been solved numerically by  iterations.
Without any doubts it provides the dynamical generation at the scalar sector. The dynamical masses of scalar and pseudoscalar bosons are degenerated which is observed
even for small coupling.
On the other hand, but  in fact  due to the mentioned $A,B$ degeneracy,  the majorana fermion remains exactly massless
\begin{equation} \label{aup}
M(p)=0\, ,
\end{equation}
which seems to be independent on the value of the coupling strength $\alpha=g^2/4\pi$. 
The equation (\ref{aup}) clearly follows from the structure of the gap equation 
(\ref{dysoni}) which  is 
 \begin{eqnarray}
M(p)\simeq \int dp M(q)[G_A(p)-G_B(p)]. 
\end{eqnarray}

The square root of selfenergy function $\Pi_{A,B}$ is displayed in the Fig. 3a. 
Recall, the results take sense only 
at the scales  where $p_E^2<\Lambda$ because of the cutoff $\Lambda$, further stress  
that the numerical results are  independent on the truncation of momentum integrals in (\ref{dysoni}) when the appropriate hard cutoff $\Lambda_H>>\Lambda$.

The observed results  obviously violates the following SUSY Ward Identities :
\begin{eqnarray} \label{ward}
\Pi_A(p)=\Pi_B(p)=Tr\frac{\not p\Sigma(p)}{4}
\end{eqnarray}
since the second equality in (\ref{ward}) is not satisfied.
One can suppose that the massless fermion 
should  be identified with the goldstino of broken supersymmetry.
On the other hand we observed the similar violation also in the explicit
massive case. To judge between the true  SUSY breaking   and  a simple effect of gap equations truncation is difficult task. 
We could learn more from the lattice simulation, which hopefully will  be performed
in  not so distinct future \cite{BONFEO2004}.

At this place we should mention the  paper \cite{BASLOR1999} where it was suggested that there is no DMG in the WZM. As the authors of  \cite{BASLOR1999} followed  the same approximations of gap equations as here, there must some weak point in their considerations since there is a clear
numerical evidence that $\Pi_A=\Pi_B\neq 0$ and here we identify this fault origin: The authors split the selfenergy into two pieces $\Pi_{A,B}(p^2)=p^2[1-F^{-1}_{A,B}(p^2)]+M_{A,B}^2(p^2)$ and they speculate that the solution $M_{A,B}^2(p^2)=0$ should follows from gap equations due to the supersymmetry. Furthermore, they speculate that the solutions of the resulting equations for two functions $F_{A,B} $ (i.e. eqs. (29) and (30) of mentioned paper) do not change the zero position of the propagator pole. However, this argument is simply wrong. Inspection by  the inverse substitution $F=\frac{p^2-M^2}{p^2-\Pi}$
we  reproduce the original gap equations for $\Pi$ (no matter what $M$ is, clearly the 'splitting' of $\Pi$ to $M,F$ is meaningless here. ) 

Furthermore, one should note that scalar propagator does not posses a real pole but it  describes resonance. Due to the behaviour of $G$ for momenta around the resonant mass it is much convenient to make a fit for selfenergy function rather them for $G$ directly. The most general dispersion  relation reads: 
\begin{eqnarray}\label{dispfit}
\Pi(p)=a(\Lambda)+b(\Lambda)p^2+\int_0^{\infty}dx\frac{\rho(x)}{x+p^2}.
\end{eqnarray}
where the function $\rho$ should be non-zero from the zero threshold. 
Clearly fitting $a,b,\rho$ we obtain the information about mass function $M(p)$ in the full
regime of momenta $p$. The procedure is limited by the efficiency of
numerical method which was minimization procedure in our case. Nevertheless 
it offers   rough numerical estimates when used in practice.

Having fitted the quantities in 
(\ref{dispfit}) we have calculated the scalar propagators at time-like regime. In principle  
the matrix element of $\psi\psi\rightarrow\psi\psi$ scattering can be read off.
In the figure 3b the typical deformed Breight-Wigner shape of  the square of  propagator  $GG^+$ is shown. In this case the coupling strength is $\alpha=0.01$ and  the obtained resonant mass is 
$m_r\simeq 2.3$ with the width $\Gamma\simeq m_r/4$. These dimensionfull variables  are scaled in  units of $10^{-5}\Lambda^2$. 
\begin{figure}
\centerline{\epsfig{figure=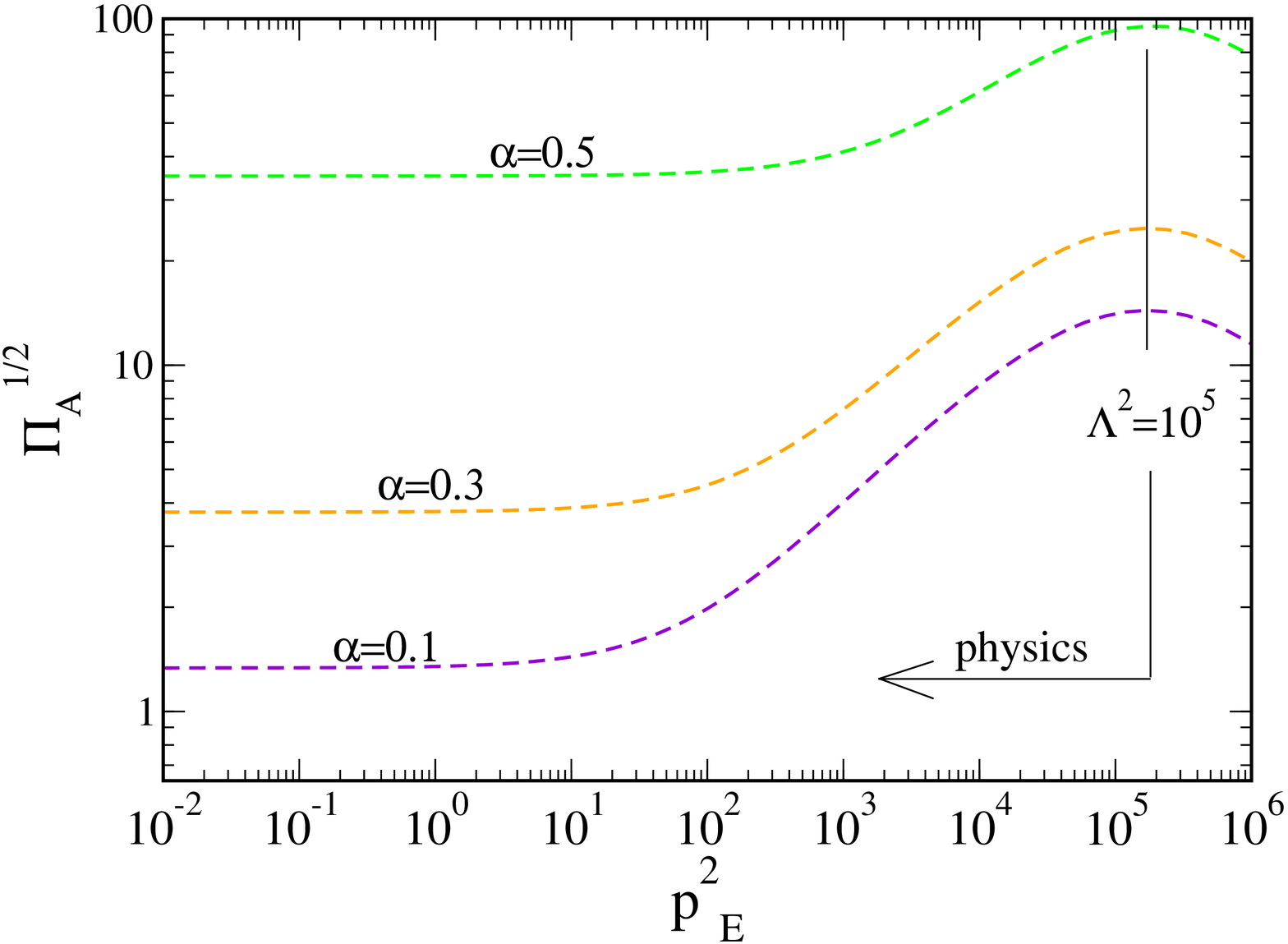,width=8truecm,height=8truecm,angle=270},
\hspace*{-1.0truecm},
\epsfig{figure=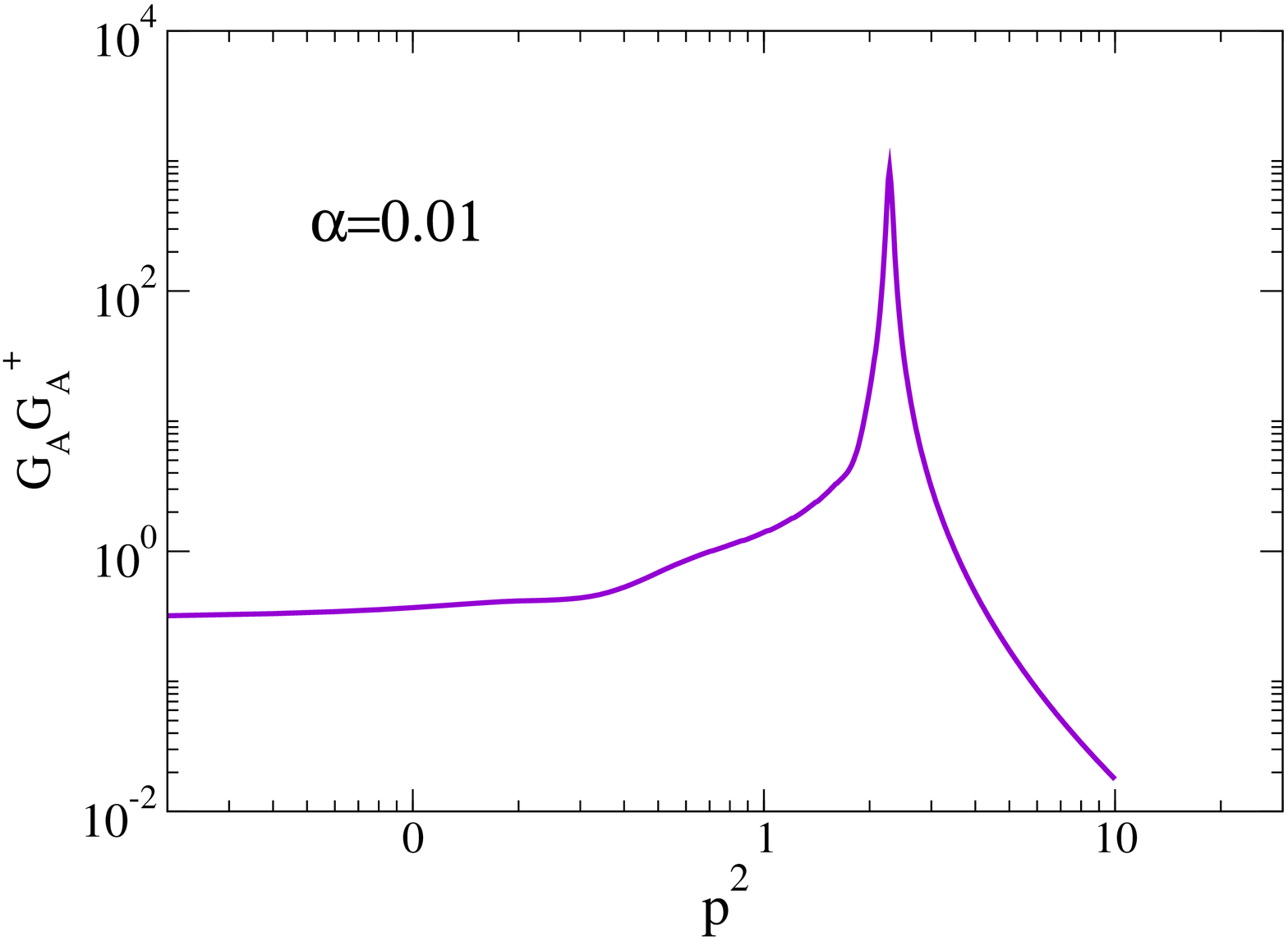,width=8truecm,height=7truecm,angle=270}}
\caption[caption]{a) Boson selfenergy b)
Scalar propagators product.}
\label{explicit2}
\end{figure}

\section{Conclusion and conjectures }

We have reviewed examples of Euclidean  quantum field theory models which provides the effect and examine the validity of standard
analytical assumptions. Especially in the case of   Yukawa theory and (strong,ladder) QED we have checked whether the known Euclidean
solutions of Schwinger-Dyson equations are consistent 
with the assumed singularity structure of the particle propagators.  
Contrary to the soft coupling case  with the explicit chiral symmetry breaking
we have found that the obtained solutions does not posses spectral (Lehmann) representation.
Although we do not recover the propagator singularity structure at all,
the observed absence of propagator pole argues that such theories 
can not describe free particles.

On the other hand, it is suggested that the  strong coupling theory may not be
in contradiction with standard (rather say then correct) analytical assumptions.
It was shown that the recent lattice data on numerical simulation for gluon propagator can be fitted by dispersion relation with continuous non-positive spectral function. Here, the proposed  absence of particle pole at real positive semi-axis of momenta is welcome
and is consistent with the confinement and unobservability of free color objects- the gluons.

The mass generation has been studied in the WZM for the first time. 
Although the majorana fermion remains massless the nontrivial solution 
for the selfenergy functions of bosons was observed. 
We have also shown the evidence for resonant behavior of processes mediated by 
the scalar $A$ and/or pseudoscalar $B$
exchanges. From our numerical investigation it  seems very likely 
that the boson propagators satisfy the  spectral representation. 

The main results of this paper is proposal of absence of 
a real particle pole in the quantum field models with strong coupling.
based on the numerical analyzes, the study presented here suggests that when fermion DMG effect appears  then the 'would be' particle modes condense or/and they are confined. 
In the certain analogy with QCD we can assume that only some composite objects -'bound states'-  can  freely move in a space. 

Of course, it is possible that some more realistic models of dynamical electroweak breaking (and hence mass generation of Standard model particles ) suffer from our simple analyzes. We have ignored particle mixing. Also the existence of soft and strong interacting sectors has not been properly modeled. The later property  is a typical requirement of the Technicolors.  
Then, if the underlying dynamics is so mysterious and  allows  free particles to receive 
their masses dynamically then there is a  white space in the road map of gap equations. 
 Otherwise we did not obtain the picture which is self-consistent with the recent sophisticated nonperturbative methods. 
One hopeful  method of such inspection
-the spectral technique-  is reviewed in this paper.

\end{document}